\documentstyle[12pt]{article}

\begin{document}

\begin{titlepage}
\setcounter{page}{1}
\title{Euclidean spinor Green's functions \\ in the
spacetime of a straight cosmic string}
\author{B. Linet \thanks{E-mail: linet.ccr.jussieu.fr} \\
\mbox{\small Laboratoire de Gravitation et Cosmologie Relativistes} \\
\mbox{\small CNRS/URA 769, Universit\'{e} Pierre et Marie Curie} \\
\mbox{\small Tour 22/12, Bo\^{\i}te Courrier 142} \\
\mbox{\small 4, Place Jussieu, 75252 PARIS CEDEX 05, France}}
\maketitle
\begin{abstract}
The spinor Green's function and the twisted
spinor Green's function in an Euclidean space with a conical-type line
singularity are generally determined. In particular, in the neighbourhood
of the point source, they are expressed as a sum of the usual Euclidean spinor
Green's function
and a regular term. In four dimension, these determinations can be used to
calculate the vacuum energy density and the twisted one for a massless spinor
field in the spacetime of a straight cosmic string. In the Minkowski
spacetime, the vacuum energy density for a massive twisted spinor
field is explicitly determined.
\newline {\em Classification PACS: 03.70.+k and 98.80.Cq.}
\end{abstract}
\end{titlepage}

\section{Introduction}

\quad \ We consider an Euclidean space with a conical-type line singularity
which is
described by the metric
\begin{equation}
\label{1.1}
ds^{2}=d\rho^{2}+B^{2}\rho^{2}d\varphi^{2}+(dx^{3})^{2}+\cdots +(dx^{n})^{2}
\end{equation}in a coordinate system ($\rho$,$\varphi$,$x^{i}$),
$i=3,\cdots ,n$ such that $\rho \geq 0$ and $0\leq \varphi <2\pi$. Metric
(\ref{1.1})
is characterized by an arbitrary positive constant $B$, different from zero.
The line
of singularity is located at $\rho =0$ but it disappears for $B=1$.
The aim of this paper is to determine in metric (\ref{1.1}) the spinor Green's
function $S^{(n)}(x,x_{0};m)$
and the twisted spinor Green's function $S^{(n)(T)}(x,x_{0};m)$.

We are motivated by the
fact that metric (\ref{1.1}) can result from the complexification of the time
coordinate $t$ in the metric
\begin{equation}
\label{1.1a}
ds^{2}=d\rho^{2}+B^{2}\rho^{2}d\varphi^{2}+dz^{2}-dt^{2}
\end{equation}
which representes a straight cosmic string \cite{vilenkin},
$B$ being related to the linear mass density $\mu$ by
$B=1-4G\mu$ (units are chosen such that $c=\hbar =1$). In this case, the
spacetime is the product of a cone by ${\cal R}^{2}$
and we have $0<B\leq 1$.
Now, to study the quantum field theory of a spinor field of mass $m$,
it is possible to
work within the Euclidean approach by performing a Wick rotation $t=-i\tau$,
 therefore the determination of the
Euclidean spinor Green's function $S_{E}(x,x_{0};m)$ and the twisted one
$S_{E}^{(T)}(x,x_{0};m)$
for the Dirac operator in metric (\ref{1.1a}) must
enable one to evaluate the vacuum expectation values of the energy-momntum
operator either for a free spinor field or for a free twisted spinor field
\cite{dewitt1,gibbons}.

Dowker \cite{dowker1} for a massless spinor field and recently Bezerra and
Bezerra de Mello
\cite{bezerra}
in three dimensions have written this spinor Green's function as contour
integrals
in the complex plan. In the present work, we give an integral expression of
$S^{(n)}(x,x_{0};m)$ and $S^{(n)(T)}(x,x_{0};m)$. We give
also a convenient form in which each spinor Green's function
is a sum of the usual spinor Green's function and a regular term, assuming that
point
$x$ is near $x_{0}$. Then the coincidence limit $x=x_{0}$ of this term
and its derivatives may enable us to calculate straightforwardly the
vacuum energy-momentum tensor.

It should be emphasized that
our method of determining $S^{(n)}(x,x_{0};m)$ and $S^{(n)(T)}(x,x_{0};m)$
will be based on the fact that they can be derived from the scalar Green's
function
$G^{(n)}_{\gamma}(x,x_{0};m)$ satisfying the following condition of periodicity
\begin{equation}
\label{1.2}
G^{(n)}_{\gamma}(\rho ,\varphi +2\pi ,x^{i};m)=\exp (2\pi i \gamma )
G^{(n)}_{\gamma}(\rho ,\varphi ,x^{i};m)
\end{equation}
for a certain constant $\gamma$ $(0\leq \gamma <1)$.
Such Green's functions have been previously derived by Guimar\~{a}es and Linet
\cite{guimaraes}.

The plan of the present work is as follows. In section 2, we recall some
properties about
the spinor fields. We determine explicitly the spinor Green's function in
section 3
and the twisted one in section 4. For massless spinor fields in
the spacetime of a straight cosmic string, the vacuum energy
density and the twisted one are calculated in section 5. The case of a massive
twisted spinor field in the Minkowski spacetime is treated in section 6. In
section 7, we add some concluding remarks.

\section{Preliminaries}

\quad \ We introduce the vierbein $e^{\mu}_{\underline{a}}$ and the Riemannian
connection
$\omega_{\mu\underline{a}\, \underline{b}}=e^{\nu}_{\underline{a}}\nabla_{\mu}
e_{\underline{b}\nu}$ where $\nabla_{\mu}$ is the covariante derivative
($\mu =1,...,n$ and $\underline{a}=1,...,n$).
We also introduce the elements $\gamma^{\underline{a}}$ which satisfy the
Clifford algebra
\begin{equation}
\label{2.1}
\gamma^{\underline{a}}\gamma^{\underline{b}}+\gamma^{\underline{b}}
\gamma^{\underline{a}}=2I\delta^{\underline{a} \, \underline{b}}
\end{equation}
where $I$ is the unit element. In the dimensions $n=2p$ and $n=2p+1$, the
$\gamma^{\underline{a}}$ are
represented by complex matrices $2^{p}\times 2^{p}$. In the present paper, we
choose these matrices so that
$\gamma^{\underline{a}\dag}=-\gamma^{\underline{a}}$
where $^{\dag}$ means Hermitian conjugate. The gauge-covariant derivative
$D_{\mu}$ of a spinor field is equal to $\partial_{\mu}+\Gamma_{\mu}$ where
\begin{equation}
\label{2.3}
\Gamma_{\mu}=\frac{1}{2}\omega_{\mu \, \underline{a} \, \underline{b}}
\sigma^{\underline{a} \, \underline{b}}
\end{equation}
with $\sigma^{\underline{a} \, \underline{b}}=(\gamma^{\underline{a}}
\gamma^{\underline{b}}-\gamma^{\underline{b}}\gamma^{\underline{a}})/4$.

In metric (\ref{1.1}), we can choose the vierbein having the components
\begin{equation}
\label{2.4}
e^{\mu}_{\underline{2}}=(0,\frac{1}{B\rho},0,\cdots ,0) \quad
e^{\mu}_{\underline{a}}=\delta^{\mu}_{\underline{a}} \quad \underline{a}\neq 2
\end{equation}
and then spinor connection (\ref{2.3}) has the components
\begin{equation}
\label{2.5}
\Gamma_{2}=-\frac{B}{4}(\gamma^{\underline{1}}\gamma^{\underline{2}}-
\gamma^{\underline{2}}\gamma^{\underline{1}}) \quad \Gamma_{i}=0 \quad i\neq 2
\end{equation}
However, choice (\ref{2.4}) of the vierbein yields some subtilities and for
instance
De Sousa Gerbert and
Deser \cite{sousa1} use another choice. The coordinates ($\rho ,\varphi$) in
metric
(\ref{1.1}) can be related to Cartesian-like coordinates ($x^{1},x^{2}$)
as explained in appendix A. In the associated
vierbein to coordinates $(x^{1},x^{2})$, the spinorial components of
a spinor field are well defined. In coordinates ($\rho ,\varphi$),
we must identify the hypersurfaces $\varphi=0$ and $\varphi=2\pi$
in the space described by metric (\ref{1.1}) but the spinorial components of
this
spinor field, expressed now in vierbein (\ref{2.4}), cannot identify.
We have to take into account the transformation law of the spinorial components
resulting from the change of vierbein.
When we use vierbein (\ref{2.4}), we have to demand that the spinorial
components $\Phi$ satisfy
\begin{equation}
\label{2.6}
\Phi (\rho ,\varphi =2\pi ,x^{i})=-\Phi (\rho ,\varphi =0,x^{i})
\end{equation}
as well as its successive derivatives with respect to $\varphi$.

We now write the Dirac operator for a spinor field of mass $m$ in metric
(\ref{1.1})
with choice (\ref{2.4}) of vierbein. The spinor Green's function
$S^{(n)}(x,x_{0};m)$
must obey the Dirac equation
\begin{equation}
\label{2.7}
(e^{\mu}_{\underline{a}}\gamma^{\underline{a}}\partial_{\mu}+\frac{\gamma^{\underline{1}}}
{2\rho}+mI)S^{(n)}=-I\frac{1}{B\rho}\delta (\rho -\rho_{0})
\delta (\varphi -\varphi_{0})\delta^{(n-2)}(x^{i}-x^{i}_{0})
\end{equation}
and, according to (\ref{2.6}), it must satisfy the requirement
\begin{equation}
\label{2.8}
S^{(n)}(\rho ,\varphi +2\pi ,x^{i};m)=-S^{(n)}(\rho ,\varphi ,x^{i};m)
\end{equation}
as well as its successive derivatives with respect to $\varphi$. Moreover, we
must impose that $S^{(n)}(x,x_{0};m)$ vanishes when the points $x$ and $x_{0}$
are infinitely separated.

\section{Determination of $S^{(n)}(x,x_{0};m)$}

\quad \ We seek $S^{(n)}(x,x_{0};m)$ in the following form
\begin{equation}
\label{3.1}
S^{(n)}(x,x_{0};m)=(e^{\mu}_{\underline{a}}\gamma^{\underline{a}}\partial_{\mu}
+\frac{\gamma^{\underline{1}}}{2\rho}-mI){\cal G}^{(n)}(x,x_{0};m)
\end{equation}
where ${\cal G}^{(n)}(x,x_{0};m)$ obey the square of the Dirac operator but
this equation
does not coincide with the Laplacian in metric (\ref{1.1}). However in the
present situation, we can put
\begin{eqnarray}
\label{3.2}
\nonumber & &{\cal G}^{(n)}(x,x_{0};m)=I\Re
H^{(n)}(x,x_{0};m)+\gamma^{\underline{1}}\gamma^{\underline{2}}
\Im H^{(n)}(x,x_{0};m)     \\
& &{\rm with} \quad H^{(n)}(x,x_{0};m)=\exp (iB\frac{\varphi -\varphi_{0}}{2})
G^{(n)}_{\gamma}(x,x_{0};m)
\end{eqnarray}
where $\Re$ and$\Im$ denote respectively the real and the imaginary part.
Then we verify that $G^{(n)}_{\gamma}(x,x_{0};m)$ satisfies the Laplacian in
metric (\ref{1.1})
\begin{eqnarray}
\label{3.3}
\nonumber & &(\frac{\partial^{2}}{\partial
\rho^{2}}+\frac{1}{\rho}\frac{\partial}{\rho}+\frac{1}{B^{2}\rho^{2}}
\frac{\partial^{2}}{\partial \varphi^{2}}+\cdots
+\frac{\partial^{2}}{\partial (x^{n})^{2}}-m^{2})G^{(n)}_{\gamma}=  \\
& &-\frac{1}{B\rho}\delta (\rho -\rho_{0})\delta (\varphi -\varphi_{0})
\delta^{(n-2)}(x^{i}-x^{i}_{0})
\end{eqnarray}
The sign of $m$ is not fixed but really we put $\mid m\mid^{2}$ in (\ref{3.3}).

With a function $G^{(n)}_{\gamma}(x,x_{0};m)$ obeying (\ref{3.3}), we
obtain a solution to equation (\ref{2.7}) but we must choose this function
so that $S^{(n)}(x,x_{0};m)$ verifies requirement (\ref{2.8}). To obtain this,
we see
that $G^{(n)}_{\gamma}(x,x_{0};m)$ must satisfy
\begin{equation}
\label{3.4}
G^{(n)}_{\gamma}(\rho ,\varphi +2\pi ,x^{i};m)=\exp (2\pi i\gamma )
G^{(n)}_{\gamma}(\rho ,\varphi ,x^{i};m)
\end{equation}
as well as its derivatives with respect to $\varphi$, where the constant
$\gamma$ is the fractional part of the number $1/2-B/2$ ($0\leq \gamma <1$).
Such scalar Green's functions
have been previously derived by Guimar\~{a}es and Linet \cite{guimaraes}.
In the usual case where metric (\ref{1.1})
describes a cone ($B\leq 1$), we have the following value of $\gamma$
\begin{equation}
\label{3.5}
\gamma =\frac{1}{2}-\frac{B}{2} \quad {\rm with}\quad 0\leq \gamma <\frac{1}{2}
\end{equation}

We had firstly derived in \cite{guimaraes} an integral expression of
$G^{(n)}_{\gamma}(x,x_{0};m)$.
As a consequence, we might obtain an integral
expression of $S^{(n)}(x,x_{0};m)$ by formulas (\ref{3.1}) and (\ref{3.2})
that we will not reproduce here. In four dimension, when the mass $m$
vanishes, we have $D^{(4)}_{\gamma}(x,x_{0})$ in closed form
\begin{equation}
\label{3.6}
D^{(4)}_{\gamma}=\frac{e^{i(\varphi -\varphi_{0})\gamma }\sinh
[\xi_{4}(1-\gamma )/B]
+e^{-i(\varphi -\varphi_{0})(1-\gamma )}\sinh (\xi_{4}\gamma /B)}
{8\pi^{2}B\rho \rho_{0}\sinh \xi_{4}[\cosh (\xi_{4} /B)-\cos (\varphi
-\varphi_{0})]}
\end{equation}
where $\xi_{4}$ is defined by
\[
\cosh
\xi_{4}=\frac{\rho^{2}+\rho^{2}_{0}+(x^{3}-x^{3}_{0})^{2}+(x^{4}-x^{4}_{0})^{2}}
{2\rho \rho_{0}}
\]
Consequently, the Euclidean spinor Green's function $S_{E}(x,x_{0})$ for the
Dirac operator
without mass can be obtained in closed form by applying formula (\ref{3.1})
with (\ref{3.2}) on expression (\ref{3.6}). With the restriction $B\leq 1$,
(\ref{3.5}) holds and we get
\begin{eqnarray}
\label{3.8}
\nonumber &
&S_{E}(x,x_{0})=(e_{\underline{a}}^{\mu}\gamma^{\underline{a}}\partial_{\mu}
+\frac{\gamma^{\underline{1}}}{2\rho}) \\
& &\nonumber \{ I\cos \frac{\varphi -\varphi_{0}}{2}
\frac{\sinh [\xi_{4}(1/B+1)/2]+\sinh [\xi_{4}(1/B-1)/2]}
{8\pi^{2}B\rho \rho_{0}\sinh \xi_{4}[\cosh \xi_{4}/B-\cos (\varphi
-\varphi_{0})]} \\
& &\gamma^{\underline{1}}\gamma^{\underline{2}}
\sin \frac{\varphi -\varphi_{0}}{2}\frac{\sinh [\xi_{4}(1/B+1)/2]-\sinh
[\xi_{4}(1/B-1)/2]}
{8\pi^{2}B\rho \rho_{0}\sinh \xi_{4}[\cosh \xi_{4}/B-\cos (\varphi
-\varphi_{0})]}\}
\end{eqnarray}
By using the identity $\exp (\psi
\gamma^{\underline{1}}\gamma^{\underline{2}})=
\cos \psi +\gamma^{\underline{1}}\gamma^{\underline{2}}\sin \psi$, we see
easily that
expression (\ref{3.8}) coincides with the result of Frolov and Serebriany
\cite{frolov}.

We had secondly derived in \cite{guimaraes} a local form of
$G^{(n)}_{\gamma}(x,x_{0};m)$ in
which it is a sum of the usual Euclidean scalar Green's function and a regular
term
$G^{*(n)}_{\gamma}(x,x_{0};m)$ valid when the points $x$ and $x_{0}$ belong
to the subset of the space defined by
\begin{equation}
\label{3.9}
\frac{\pi}{B}-2\pi <\varphi -\varphi_{0}<2\pi -\frac{\pi}{B}
\end{equation}
in the case $B>1/2$ in which we restrict ourselves. We have
\begin{equation}
\label{3.10}
G^{(n)}_{\gamma}=\frac{m^{n/2-1}}{(2\pi )^{n/2}r^{n/2-1}_{n}}K_{n/2-1}(mr_{n})+
G^{*(n)}_{\gamma}(x,x_{0};m)
\end{equation}
with $r_{n}=\sqrt{\rho^{2}+\rho_{0}^{2}+2\rho \rho_{0}\cos B(\varphi
-\varphi_{0})
+\cdots +(x^{n}-x_{0}^{n})^{2}}$, the $K_{\mu}$ being the modified Bessel
function,
and the regular term $G_{\gamma}^{*(n)}(x,x_{0};m)$ is given by
\begin{equation}
\label{3.10a}
G_{\gamma}^{*(n)}=\frac{m^{n/2-1}}{(2\pi)^{n/2+1}B}\int_{0}^{\infty}
\frac{K_{n/2-1}[mR_{n}(u)]}{[R_{n}(u)]^{n/2-1}}F_{B}^{(\gamma )}(u,\varphi
-\varphi_{0})du
\end{equation}
with $R_{n}(u)=\sqrt{\rho^{2}+\rho_{0}^{2}+2\rho \rho_{0}\cosh u+\cdots
+(x^{n}-x_{0}^{n})^{2}}$
and the function $F_{B}^{(\gamma )}(u,\psi )$ has the expression
\begin{eqnarray}
\label{3.10b}
\nonumber & &F_{B}^{(\gamma )}(u,\psi )=i\frac{e^{i(\psi +\pi /B)\gamma} \cosh
[\frac{u(1-\gamma )}{B}]
-e^{-i(\psi +\pi /B)(1-\gamma )}\cosh \frac{u\gamma}{B}}{\cosh \frac{u}{B}-\cos
(\psi
+\frac{\pi}{B})} \\
& &-i\frac{e^{i(\psi -\pi /B)\gamma} \cosh [\frac{u(1-\gamma}{B}]
-e^{-i(\psi -\pi /B)(1-\gamma )}\cosh \frac{u\gamma}{B}}{\cosh \frac{u}{B}
-\cos (\psi -\frac{\pi}{B})}
\end{eqnarray}
{}From formulas (\ref{3.1}) and (\ref{3.2}), we can determine the
spinor Green's function $S^{(n)}(x,x_{0};m)$ under a local form valid in the
subset (\ref{3.9}) of the space. We obtain
\begin{eqnarray}
\label{3.11}
\nonumber &
&S^{(n)}=(e_{\underline{a}}^{\mu}\gamma^{\underline{a}}\partial_{\mu}+
\frac{\gamma^{\underline{1}}}{2\rho}-mI) \\
\nonumber & &\{ (I\cos B\frac{\varphi
-\varphi_{0}}{2}+\gamma^{\underline{1}}\gamma^{\underline{2}}
\sin B\frac{\varphi -\varphi_{0}}{2})\frac{m^{n/2-1}}{(2\pi
)^{n/2}r^{n/2-1}_{n}}
K_{n/2-1}(mr_{n})\}    \\
& &+S^{*(n)}
\end{eqnarray}
where the regular term $S^{*(n)}(x,x_{0};m)$ can be expressed by formula
(\ref{3.1})
\begin{eqnarray}
\label{3.12}
\nonumber & &S^{*(n)}(x,x_{0};m)=(e_{\underline{a}}^{\mu}\gamma^{\underline{a}}
\partial_{\mu}+\frac{\gamma^{\underline{1}}}{2\rho}-mI) \\
& &\{ I\Re H^{*(n)}(x,x_{0};m)+\gamma^{\underline{1}}
\gamma^{\underline{2}}\Im H^{*(n)}(x,x_{0};m) \}
\end{eqnarray}
The regular term $H^{*(n)}(x,x_{0};m)$ appearing in (\ref{3.12}) is given by
\begin{equation}
\label{3.13}
H^{*(n)}=\frac{m^{n/2-1}}{(2\pi )^{n/2+1}B}\int_{0}^{\infty}
\frac{K_{n/2-1}[mR_{n}(u)]}{[R_{n}(u)]^{n/2-1}}
H^{(\gamma )}_{B}(u,\varphi -\varphi_{0})du
\end{equation}
where the function $H^{(\gamma )}_{B}(u,\psi )$ has the expression
\begin{equation}
\label{3.14}
H_{B}^{(\gamma )}(u,\psi )=\exp (iB\frac{\psi}{2})F_{B}^{(\gamma )}(u,\psi )
\end{equation}
Local form (\ref{3.11}) is valid for $B>1/2$.

Assuming moreover $B\leq 1$, we have $\gamma$ given by (\ref{3.5}). We
specialize expression (\ref{3.14}) for this value of $\gamma$; we find the real
part
\begin{eqnarray}
\label{3.15}
\nonumber & &\Re H_{B}^{(\frac{1}{2}-\frac{B}{2})}(u,\psi )=\cos
(\frac{\psi}{2}+\frac{\pi}{2B})\frac{\cosh [\frac{u}{2}
(\frac{1}{B}+1)]-\cosh [\frac{u}{2}(\frac{1}{B}-1)]}
{\cosh \frac{u}{B}-\cos (\psi +\frac{\pi}{B})} \\
& &+\cos (\frac{\psi}{2}-\frac{\pi}{2B})\frac{\cosh
[\frac{u}{2}(\frac{1}{B}+1)]
-\cosh [\frac{u}{2}(\frac{1}{B}-1)]}{\cosh \frac{u}{B}-\cos (\psi -
\frac{\pi}{B})}
\end{eqnarray}
and the imaginary part
\begin{eqnarray}
\label{3.16}
\nonumber & &\Im H_{B}^{(\frac{1}{2}-\frac{B}{2})}(u,\psi )=\sin
(\frac{\psi}{2}+\frac{\pi}{2B})\frac{\cosh [\frac{u}{2}
(\frac{1}{B}+1)]+\cosh [\frac{u}{2}(\frac{1}{B}-1)]}{\cosh \frac{u}{B}-
\cos (\psi +\frac{\pi}{B})} \\
& &+\sin (\frac{\psi}{2}-\frac{\pi}{2B})\frac{\cosh
[\frac{u}{2}(\frac{1}{B}+1)]
+\cosh [\frac{u}{2}(\frac{1}{B}-1)]}{\cosh \frac{u}{B}-
\cos (\psi -\frac{\pi}{B})}
\end{eqnarray}

By setting $n=4$ in formula (\ref{3.11}), we obtain
the expression of the Euclidean
spinor Green's function $S_{E}(x,x_{0};m)$ for
point $x$ near $x_{0}$ in the case $B>1/2$. Since it will be needed in a next
section, we write down it when the mass $m$ vanishes; we have
\begin{eqnarray}
\label{3.17}
\nonumber &
&S_{E}=(e_{\underline{a}}^{\mu}\gamma^{\underline{a}}\partial_{\mu}+
\frac{\gamma^{\underline{1}}}{2\rho}) \\
\nonumber & &\{ (I\cos B\frac{\varphi
-\varphi_{0}}{2}+\gamma^{\underline{1}}\gamma^{\underline{2}}
\sin B\frac{\varphi -\varphi_{0}}{2})\frac{1}{4\pi^{2}r_{4}^{2}} \\
& &+I\Re H^{*(4)}+\gamma^{\underline{1}}\gamma^{\underline{2}}\Im H^{*(4)} \}
\end{eqnarray}
where the regular term $H^{*(4)}(x,x_{0};m)$ has now the integral expression
\begin{equation}
\label{3.18}
H^{*(4)}=\frac{1}{8\pi^{3}B}\int_{0}^{\infty}\frac{1}{[R_{4}(u)]^{2}}
H_{B}^{(\gamma )}(u,\varphi -\varphi_{0})du
\end{equation}

\section{Determination of $S^{(n)(T)}(x,x_{0};m)$}

\quad \ Isham \cite{isham} has shown that twisted fields can be defined
in a spacetime which is not simply connected. In space described by metric
(\ref{1.1}), the axis $\rho =0$ can
be removed from the space. So, the twisted fields are obtained by requiring
that
they are antiperiodic about this axis. Twisted scalar fields have already
considered in the spacetime of a straight cosmic string \cite{smith} and in
the particular case of the Minkowski spacetime \cite{ford,serebriany}.

In metric (\ref{1.1}), the property of antiperiodicity for a
twisted spinor field is required
in the Cartesian-like coordinates $(x^{1},x^{2})$ with the associated vierbein.
Now
we work with choice (\ref{2.4}) of vierbein, therefore we demand that the
spinorial components of the twisted spinor field $\Phi$ satisfy
\begin{equation}
\label{5.1}
\Phi (\rho ,\varphi =2\pi ,x^{i};m)=\Phi (\rho ,\varphi =0,x^{i};m)
\end{equation}
instead of requirement (\ref{2.6}) for a spinor field. The twisted spinor
Green's function $S^{(n)(T)}(x,x_{0};m)$ for the Dirac operator must satisfy
the condition
\begin{equation}
\label{5.2}
S^{(n)(T)}(\rho ,\varphi +2\pi ,x^{i};m)=S^{(n)(T)}(\rho ,\varphi ,x^{i};m)
\end{equation}
as well as its sucessive derivatives with respect to $\varphi$. It is obvious
that
$S^{(n)(T)}(x,x_{0};m)$ can be determined by formulas (\ref{3.1})
and (\ref{3.2}) following the method developped in section 3 but the scalar
Green's function $G_{\gamma}^{(n)}(x,x_{0};m)$ is now characterized by a
constant
$\gamma^{(T)}$ which is the fractional part of the number $1-B/2$. In the case
where $B\leq 1$, we have the following value of $\gamma^{(T)}$
\begin{equation}
\label{5.3}
\gamma^{(T)}=1-\frac{B}{2} \quad {\rm with } \quad \frac{1}{2}\leq
\gamma^{(T)}<1
\end{equation}

In metric (\ref{1.1a}) of the spacetime of a straight cosmic string, the
quantum field theory for a twisted spinor field can be done from the
Euclidean twisted spinor Green's function $S_{E}^{(T)}(x,x_{0};m)$ obtained
by setting $n=4$. By this way, we determine an explicit expression of
the Euclidean twisted spinor Green's function in the massless case. In the
massive
case, we can obtain $S_{E}^{(T)}(x,x_{0};m)$ in a local form valid in subset
(\ref{3.9})
in which we take $\gamma^{(T)}$ instead of $\gamma$ in formula (\ref{3.12}).

\section{Vacuum energy density (massless spinor)}

\quad \ Within the Euclidean quantum field theory of a spinor field in the
spacetime
of a straight cosmic string, the fondamental quantity in the free case is the
Euclidean spinor Green's function $S_{E}(x,x_{0};m)$ which coincides with
$S^{(4)}(x,x_{0};m)$. To renormalize, it is
convenient to use local form (\ref{3.11}) of the spinor Green's function valid
for $B>1/2$. The expectation values of the energy-momentum operator is
performed
by removing the usual spinor Euclidean Green's function in expression
(\ref{3.11}). Since metric
(\ref{1.1a}) is locally flat we obtain merely the vacuum energy-momentum
tensor in the Euclidean approach by the usual formula  \cite{christensen}
\begin{equation}
\label{4.3}
<T_{\mu \nu}>=\frac{1}{4}tr[\gamma^{\underline{a}}(e_{\underline{a}\mu}
(\partial_{\nu}-\partial_{\nu_{0}})+e_{\underline{a}\nu}(\partial{\mu}-
\partial_{\mu_{0}})S^{*(4)}(x,x_{0};m)]\mid_{x=x_{0}}
\end{equation}
where $\partial_{\nu_{0}}$ denotes the derivative with respect to $x_{0}$. We
point out
that only the antisymmetrical part of $S^{*(4)}(x,x_{0};m)$ with respect
to $x$ and $x_{0}$ occurs in formula (\ref{4.3}). The vacuum energy-momentum
tensor
is conserved.

The vacuum energy density $<T_{tt}>$ is given by $-<T_{\tau \tau}>$ since
$t=-i\tau$ in a Wick rotation. From expression (\ref{4.3}), we obtain the
general form
\begin{equation}
\label{4.2}
<T_{tt}(x)>=-4\partial_{\tau \tau}\Re H^{*(4)}(x,x_{0};m)\mid_{x=x_{0}}
\end{equation}
where we have used the fact that expression (\ref{3.13}) of
$H^{*(4)}(x,x_{0};m)$
depends on $\tau$ and $\tau_{0}$ by $(\tau -\tau_{0})^{2}$.

We now confine ourselves to massless spinors.
According to expression (\ref{3.13}) of $H^{*(4)}(x,x_{0})$, we obtain
immediately
\begin{equation}
\label{4.3a}
<T_{tt} (x)>=\frac{1}{\pi^{3}B}\int_{0}^{\infty}\frac{1}{[R_{4}(u)]^{4}}
H_{B}^{(\gamma )}(u,0)du
\end{equation}
We rewrite (\ref{4.3a}) in the form
\begin{equation}
\label{4.4}
<T_{tt}(x)>=\frac{2}{\rho^{4}}w_{4}(\gamma )
\end{equation}
where $w_{4}(\gamma )$ is the following integral
\begin{equation}
\label{4.5}
w_{4}(\gamma )=-\frac{1}{4\pi^{3}B}\int_{0}^{\infty}
\frac{\sin \frac{\pi \gamma}{B}\cosh [\frac{u(1-\gamma )}{B}]
+\sin [\frac{\pi (1-\gamma )}{B}]\cosh \frac{u\gamma}{B}}
{(1+\cosh u)^{2}(\cosh \frac{u}{B}-\cos \frac{\pi}{B})}
\end{equation}
Integrals of type (\ref{4.5}) have been studied by Dowker
\cite{dowker1,dowker2}. For $B>1/2$, we have
the explicit expression
\begin{equation}
\label{4.6}
w_{4}(\gamma )=-\frac{1}{720\pi^{2}}\{ 11-\frac{15}{B^{2}}[4(\gamma
-\frac{1}{2})^{2}
-\frac{1}{3}]+\frac{15}{8B^{4}}[16(\gamma -\frac{1}{2})^{4}-8(\gamma
-\frac{1}{2})^{2}
+\frac{7}{15}] \}
\end{equation}
So, for $B>1/2$, we have explicitly determined in general the vacuum energy
density
which is given by
by formula (\ref{4.4}). Its value depends on the choice of $\gamma$.
With restriction $B\leq 1$, $\gamma$ is given by (\ref{3.5})
and then we find the result of Frolov and Serebriany \cite{frolov}
\begin{equation}
\label{4.7}
<T_{tt}(x)>=-\frac{1}{2880\pi^{2}\rho^{4}}[(\frac{1}{B^{2}}-1)(\frac{7}{B^{2}}+17)]
\end{equation}
It vanishes for $B=1$.

We now turn to the twisted case where $\gamma$
is given by (\ref{5.3}) when $B\leq 1$. From formula (\ref{4.4}) we find  the
twisted vacuum energy density
\begin{equation}
\label{4.8}
<T_{tt}(x)>^{(T)}=-\frac{1}{360\pi^{2}\rho^{4}}[-\frac{17}{8}+\frac{45}{2B}-
\frac{5}{2B^{2}}-\frac{1}{B^{4}}]
\end{equation}
As noticed by DeWitt et al \cite{dewitt2}, the twisted vacuum has the lower
vacuum
energy density in the case of spinor fields. Actually, we see that
\begin{equation}
\label{4.9}
<T_{tt}(x)>^{(T)}< \, <T_{tt}(x)> \quad {\rm for } \quad \frac{1}{2}<B\leq 1
\end{equation}
in our situation.
They coincide in the limit where the constant $B=1/2$.

More generally for a massive spinor field,
our method should be enabled us to express the vacuum
energy density under an explicit integral form.

\section{Massive twisted spinor field in Minkowski spacetime}

\quad \ In the Minkowski spacetime, characterized by $B=1$, a twisted spinor
field can be defined.
The massless case has been already considered by Ford \cite{ford}. The method
of the present
paper can be applied to evaluate the twisted vacuum energy density.
Since $\gamma^{(T)}=1/2$ in this case,
the twisted spinor Green's function $S_{E}^{(T)}(x,x_{0};m)$
is derived from the twisted scalar Green's function $G_{1/2}^{(4)}(x,x_{0};m)$.

For $B>1/2$, we write down $S_{E}^{(T)}(x,x_{0};m)$ in a local form
valid in subset (\ref{3.9}) of the Euclidean space
\begin{eqnarray}
\label{6.1}
\nonumber &
&S_{E}^{(T)}=(e_{\underline{a}}^{\mu}\gamma^{\underline{a}}\partial_{\mu}
+\frac{\gamma^{\underline{1}}}{2\rho}-mI) \\
\nonumber & &\{ \cos \frac{\varphi
-\varphi_{0}}{2}+\gamma^{\underline{1}}\gamma^{\underline{2}}
\sin \frac{\varphi -\varphi_{0}}{2})\frac{m}{4\pi^{2}r_{4}}K_{1}(mr_{4}) \\
& &I\Re H^{*(4)}+\gamma^{\underline{1}}\gamma^{\underline{2}}\Im H^{*(4)} \}
\end{eqnarray}
where the regular term $H^{*(4)}(x,x_{0};m)$ has now the integral expression
\begin{equation}
\label{6.2}
H^{*(4)}=\frac{m}{8\pi^{3}}\int_{0}^{\infty}\frac{K_{1}[mR_{4}(u)]}{R_{4}(u)}
H_{1}^{(1/2)}(u,\varphi -\varphi_{0})du
\end{equation}
in which $H_{1}^{(1/2)}(u,\psi )$ is obtained from (\ref{3.14}) under the form
\begin{equation}
\label{6.3}
H_{1}^{(1/2)}(u,\psi )=-2(1+\cos \psi +i\sin \psi )\frac{\cosh u/2}{\cosh
u+\cos \psi}
\end{equation}

By using formula (\ref{4.2}), we obtain the twisted vacuum energy density
as a definite integral
\begin{eqnarray}
\label{6.4}
\nonumber & &<T_{tt}(x)>^{(T)}=-\frac{m^{2}}{2\pi^{3}\rho^{2}}\int_{0}^{\infty}
\frac{K_{0}(2m\rho \cosh v)}{(\cosh v)^{3}}dv \\
& &-\frac{m}{2\pi^{3}\rho^{3}}\int_{0}^{\infty}
\frac{K_{1}(2m\rho \cosh v)}{(\cosh v)^{4}}dv
\end{eqnarray}
In the appendix B, we give some definite integrals of this type which are
useful
for our problem. We can thereby obtain the expression of the twisted vacuum
energy density in closed form
\begin{eqnarray}
\label{6.5}
\nonumber &
&<T_{tt}(x)>^{(T)}=-\frac{m^{2}}{32\pi^{2}\rho^{2}}[4m^{2}\rho^{2}E_{1}(2m\rho
) \\
& &+(1-2m\rho +\frac{3}{m\rho}+\frac{3}{2m^{2}\rho^{2}})\exp (-2m\rho )]
\end{eqnarray}
where $E_{1}$ is the exponential integral function. The twisted vacuum energy
density is always negative and its range is $1/2m$.

In the limit where $m\rho \ll 1$, expression (\ref{6.5}) becomes asymptotically
\begin{equation}
\label{6.6}
<T_{tt}(x)>^{(T)}\sim -\frac{3}{64\pi^{2}\rho^{4}}
\end{equation}
The twisted vacuum energy density (\ref{6.6}) coincides with expression
(\ref{4.8}),
directly established for a massless twisted spinor field, in which we set
$B=1$.

\section{Conclusion}

\quad \ We have explicitly determined the spinor Green's function and the
twisted spinor Green's
function in an Euclidean space with a conical-type line
singularity. It should be emphasized that their local expression
when point $x$ is near $x_{0}$
is always convenient to evaluate straightforwardly the
vacuum energy-momentum tensor within the framework of quantum field theory of
free spinor fields.
In the massless case, we have calculated the vacuum energy density
and the twisted vacuum energy density for an arbitrary
constant $B$ $(B>1/2)$ in the four dimensional spacetime. In the massive case,
we might obtain it under the form of definite integrals.
For a massive twisted spinor field in the Minkowski spacetime,
we have performed the integration in terms of elementary functions.

Our method can also give the spinor Green's functions,
in the vierbein used in this work, which satisfy the condition
\[
S^{(n)}_{\Phi /\Phi_{0}}(\rho ,\varphi +2\pi ,x^{i})=-\exp (2\pi i\Phi
/\Phi_{0})
S^{(n)}_{\Phi /\Phi_{0}}(\rho ,\varphi ,x^{i})
\]
for constants $\Phi /\Phi_{0}$. We must simply introduce in our formulas the
scalar Green's function
$G^{(n)}_{\gamma}(x,x_{0};m)$ where $\gamma$ is the fractional part of the
number
$1/2-B/2-\Phi /\Phi_{0}$. In contrast to the case of a scalar charged field,
the
quantum field theory of a charged spinor field in a magnetic flux $\Phi$,
$\Phi_{0}$
being the quantum flux, cannot describe by the $S^{(4)}_{\Phi
/\Phi_{0}}(x,x_{0};m)$
determined in this manner.
Indeed, the axis $\rho =0$ is excluded in this kind of method \cite{hagen}. The
boundary conditions
on the axis $\rho =0$ must be carefully examined in this situation
\cite{sousa2,coutinho}.

\appendix

\section{Appendix: transformation law of the spinorial components}

\quad \ By making the coordinate transformation $\overline{\rho}=\rho^{1/B}$,
metric (\ref{1.1}) takes the form
\[
ds^{2}=B^{2}\overline{\rho}^{2B-2}(d\overline{\rho}^{2}+\overline{\rho}^{2}
d\varphi^{2})+(dx^{3})^{2}+\cdots (dx^{n})^{2}
\]
In the coordinates $(\overline{\rho},
\varphi ,x^{i})$, vierbein (\ref{2.4}) has now the components
\[
e_{\underline{1}}^{\mu}=(\frac{1}{B\overline{\rho}^{B-1}}, 0,0,...,0)
\]
\[
e_{\underline{2}}^{\mu}=(0,\frac{1}{B\overline{\rho}^{B}},0,...,0)
\]
\[
e_{\underline{a}}^{\mu}=\delta_{\underline{a}}^{\mu} \quad
\underline{a}=3,...,n
\]
We now introduce the Cartesian coordinates $(x^{1},x^{2})$ related to
$(\overline{\rho},
\varphi )$. The vierbein $\overline{e}_{\underline{a}}^{\mu}$ associated to the
Cartesian coordinates $(x^{\mu})$ have the components
\[
\overline{e}_{\underline{1}}^{\mu}=(\frac{1}{B\overline{\rho}^{B-1}},0,0,...,0)
\]
\[
\overline{e}_{\underline{2}}^{\mu}=(0,\frac{1}{B\overline{\rho}^{B-1}},0,...,0)
\]
\[
\overline{e}_{\underline{a}}^{\mu}=\delta_{\underline{a}}
^{\mu} \quad \underline{a}=3,...,n
\]
Let a spinor well defined in the coordinates $(x^{\mu})$; its spinorial
components $\overline{\Phi}$, defined relatively to the associated vierbein
$\overline{e}_{\underline{a}}^{\mu}$, are regular.
But in vierbein $e_{\underline{a}}^{\mu}$, this spinor has the following
spinorial components $\Phi$
\[
\Phi (\overline{\rho},\varphi ,x^{i})=(I\cos \frac{\varphi}{2}+
\gamma^{\underline{1}}\gamma^{\underline{2}}\sin \frac{\varphi}{2})
\overline{\Phi}(\overline{\rho},\varphi ,x^{i})
\]
that we have expressed in coordinates $(\overline{\rho},\varphi ,x^{i})$.
We can immediately return to usual coordinates $(\rho ,\varphi ,x^{i})$ and
we see that we obtain requirement (\ref{2.6}) for the spinorial components
$\Phi$.

The spinor Green's function is a bi-spinor whose the spinorial components
$\overline{S}^{(n)}(x,x_{0};m)$ defined relatively to vierbein
$\overline{e}_{\underline{a}}^{\mu}$ are well
defined. The spinorial components in vierbein $e_{\underline{a}}^{\mu}$
are obtained by a transformation law which can be recast in the form
\[
S^{(n)}(\overline{\rho},\varphi ,x^{i})=(I\cos \frac{\varphi -\varphi_{0}}{2}
+\gamma^{\underline{1}}\gamma^{\underline{2}}\sin \frac{\varphi
-\varphi_{0}}{2})
\overline{S}^{(n)}(\overline{\rho},\varphi ,x^{i})
\]
It yields also condition (\ref{2.8}).

\section{Appendix: a list of some definite integrals}

\quad \ To our knowledge, these following definite integrals are not in the
mathematical tables
under this form. The two first integrals have been derived in the context of
a twisted scalar field in the spacetime of a straight cosmic string
\cite{linet}.
For a positive real number $b$, we have the formulas
\[
\int_{0}^{\infty}\frac{K_{0}(b\cosh v)}{\cosh v}dv=\frac{\pi}{2}E_{1}(b)
\]
\[
\int_{0}^{\infty}\frac{K_{1}(b\cosh v)}{(\cosh v)^{2}}dv=\frac{\pi}{4}[
-bE_{1}(b)+(\frac{1}{b}+1)\exp (-b)]
\]
in which $K_{\mu}$ denotes the modified Bessel function and $E_{1}$ the
exponential integral function.

The two other integrals can be obtained by performing some integrations by
parts and by using the above mentioned integrals. Furthermore, we recall
the identities
\[
K_{0}'(x)=-K_{1}(x) \quad {\rm and } \quad K_{1}'(x)=-K_{0}-\frac{1}{x}K_{1}(x)
\]
We give the final results without proof
\[
\int_{0}^{\infty}\frac{K_{0}(b\cosh v)}{(\cosh v)^{3}}dv=\frac{\pi}{8}[
(2+b^{2})E_{1}(b)+(1-b)\exp (-b)]
\]
\[
\int_{0}^{\infty}\frac{K_{1}(b\cosh v)}{(\cosh v)^{4}}dv=\frac{\pi}{32}[
(-4b-b^{3})E_{1}(b)
+(\frac{6}{b}+6-b+b^{2})\exp (-b)]
\]

We have checked these definite integrals by a numerical analysis.


\newpage

\begin{thebibliography}{99}

\bibitem{vilenkin} A. Vilenkin, in {\em Three hundred years of gravitation},
S. W. Hawking and W. Israel (eds.) (Cambridge University Press, Cambridge,
1987)
p. 499.
\bibitem{dewitt1} B. S. DeWitt, in {\em Relativity, groups and topology II},
 B. S. DeWitt and R. Stora (eds.) (North-Holland, Amsterdam, 1984) p. 381.
\bibitem{gibbons} G. W. Gibbons, in {\em General relativity: an Einstein
centenary
survey}, S. W. Hawking and W. Israel (eds.) (Cambridge University Press,
Cambridge, 1979) p. 639.
\bibitem{dowker1} J. S. Dowker, Phys. Rev. D {\bf 36}, 3742 (1987).
\bibitem{bezerra} V. B. Bezerra and E. R. Bezerra de Mello, Class. Quant. Grav.
{\bf 11}, 457 (1994).
\bibitem{guimaraes} M. E. X. Guimar\~{a}es and B. Linet, Comm. Math. Phys.
{\bf 165}, 297 (1994).
\bibitem{sousa1} P. De Sousa Gerbert and R. Jackiw, Comm. Math. Phys. {\bf
124},
229 (1989).
\bibitem{isham} C. J. Isham, Proc. Roy. Soc. A {\bf 364}, 591 (1978).
\bibitem{christensen} S. M. Christensen, Phys. Rev. D {\bf 17}, 946 (1978).
\bibitem{dowker2} J. S. Dowker, Phys. Rev. D {\bf 36}, 3095 (1987).
\bibitem{frolov} V. P. Frolov and E. M. Serebriany, Phys. Rev. D {\bf 35},
3779 (1987).
\bibitem{smith} A. G. Smith, in {\em The formation and evolution of cosmic
strings}, G. W. Gibbons, S. W. Hawking and T. Vaschapati (eds) (Cambridge
University Press, Cambridge, 1990) p. 262.
\bibitem{ford} L. H. Ford, Phys. Rev. D {\bf 21}, 949 (1980).
\bibitem{serebriany} E. M. Serebriany, Theor. Math. Phys. {\bf 64}, 846 (1985).
\bibitem{dewitt2} B. S. DeWitt, C. F. Hart and C. J. Isham, Physica A {\bf 96},
197 (1979).
\bibitem{hagen} C. R. Hagen, Phys. Rev. Lett. {\bf 64}, 503 (1990).
\bibitem{sousa2} P. De Sousa Gerbert, Phys. Rev. D {\bf 40}, 1346 (1989).
\bibitem{coutinho} F. A. B. Coutinho and J. Fernando Perez, Phys. Rev. D
{\bf 48}, 932 (1993).
\bibitem{linet} B. Linet, Int. J. Mod. Phys. D {\bf 1}, 371 (1992).

\end{thebibliography}
\end{document}